\documentstyle[12pt]{article}
\advance\textheight by \topskip
\def \sl  { / {\hskip - 0.27 cm  {\bf P}}}

\begin{document}

\begin{flushright}
SU-ITP-97-26\\

hep-th/9705118\\
May 15, 1997
\end{flushright}
\vspace{.5cm}

\begin{center}
\baselineskip=16pt

{\Large\bf   Volkov-Akulov  theory and D-branes}\footnote{Contribution to
Supersymmetry and Quantum Field
Theory, International Seminar dedicated to the memory of D.V. Volkov,
Kharkov State University (Kharkov, Ukraine) January 5-7, 1997.}
 \\

\

\vskip 1 cm

{\bf  Renata Kallosh \footnote{This work   is
supported by the NSF grant PHY-9219345.
}}

\vskip 1cm

{\em Physics Department, Stanford
University, Stanford, CA 94305-4060, USA\\
kallosh@physics.stanford.edu
}

\end{center}

\vskip 1 cm
\centerline{\bf ABSTRACT}
\vspace{-0.3cm}
\begin{quote}
 The action of supersymmetric Born-Infeld theory  (D-9-brane in a Lorentz
covariant static gauge)
has a geometric form of the  Volkov-Akulov-type. The  first  non-linearly
realized supersymmetry can be made manifest, the second  world-volume
supersymmetry is not manifest.  We also study the analogous  2 supersymmetries
of the quadratic action of the covariantly quantized D-0-brane. We show that
the Hamiltonian and the BRST operator are build from these two
supersymmetry generators.

\end{quote}
\normalsize

\baselineskip=16pt

\noindent This article is dedicated to the memory of D. V. Volkov whose
insights into the nature of supersymmetry and geometry proved to be
enlightening for few generations of high-energy physicists. His ideas inspired
the most active developments in theoretical physics over the last quarter of
the century.

\vskip 1 cm

Extended objects with global supersymmetry have local $\kappa$-symmetry. This
symmetry is difficult to quantize in Lorentz covariant gauges keeping finite
number of fields in the theory. A revival of
interest to $\kappa$-symmetric objects is due to the recent discovery of
D-p-branes \cite{Pol},  $\kappa$-symmetric non-linear effective actions and/or
equations of
motion for D-branes \cite{Ce1,Ag1,Be1,Ho1,Hull, BKOP}
 and the M-5-brane action
\cite{Ho2,Ba1,Ag2}, complementing the $\kappa$-symmetric superstring
and M-2-brane \cite{grsh,begpkt}.  A nice review on worldvolume actions in a
doubly supersymmetric geometric approach initiated by D. V. Volkov   is
presented in these Proceedings \cite{BPST}.

The new issues in quantization of D-branes have been analyzed recently in
\cite{Ag1, BKOP,K97}.

The $\kappa$-symmetric D-brane action in the flat background
geometry\footnote{We use
notation of \cite{Ag1}.} consists of the
Born-Infeld-Nambu-Goto term $S_1$ and Wess-Zumino term $S_2$:
\begin{equation}
S_{\rm DBI} +S_{\rm WZ} =T\left (  - \int d^{p + 1} \sigma\, \sqrt{- {\rm {\rm
det}} (G_{\mu\nu} + {\cal F}_{\mu\nu})} +\int \Omega_{p + 1} \right) \ .
\label{action}\end{equation}
Here $T$ is the tension of the D-brane, $G_{\mu\nu}$ is the  manifestly
supersymmetric induced world-volume metric
\begin{equation}
G_{\mu\nu} = \eta_{mn} \Pi_\mu^m \Pi_\nu^n \ , \qquad
\Pi_\mu^m = \partial_\mu X^m - \bar\theta \Gamma^m \partial_\mu \theta \ ,
\end{equation}
and ${\cal F}_{\mu\nu}$ is a manifestly supersymmetric Born-Infeld field
strength (for $p$ even)
\footnote{We define
 spinors  for even $p$ as $\theta = \theta_1 + \theta_2$ where  $\theta_1
\equiv
{1\over
2}(1+\Gamma_{11})\theta$
 and $\theta_2 \equiv {1\over 2}(1-\Gamma_{11})\theta$.}

\begin{equation}
{\cal F}_{\mu\nu} \equiv F_{\mu\nu} - b_{\mu\nu} =  \Bigl[\partial_\mu A_\nu -
\bar\theta \Gamma_{11} \Gamma_m \partial_{\mu}\theta
\Bigl(\partial_{\nu} X^m - {1\over 2} \bar
\theta \Gamma^m \partial_{\nu}\theta\Bigr)\Bigr] - (\mu \leftrightarrow \nu) \
{}.
\end{equation}
 When $p$ is odd, $\Gamma_{11}$ is replaced by
$\tau_3\otimes I$.  The action has
global supersymmetry
\begin{equation}
\delta_\epsilon \theta = \epsilon, \qquad \delta_\epsilon X^m = \bar\epsilon
\Gamma^m
\theta \ . \label{susytrans}
\end{equation}
and local $\kappa$-supersymmetry:
\begin{equation}
\delta X^m = \bar\theta \Gamma^m \delta\theta = - \delta \bar\theta \Gamma^m
\theta,
\label{Xtrans} \qquad
\delta \bar\theta = \bar \kappa (1+\Gamma),
\end{equation}
and \begin{equation}
\label{rotGamma}
\Gamma = e^{a\over 2} \Gamma'_{(0)} e^{-{a\over 2}}\, , \qquad
a=\cases{+{1\over2} Y_{jk} \gamma^{jk} \Gamma_{11}\qquad\qquad &{\rm
IIA\, ,}\cr
-{1\over2} Y_{jk}   \gamma^{jk} \sigma_3\otimes 1\qquad\qquad &{\rm IIB\, .}}
\end{equation}

Here $\Gamma_{(0)}'$ is the product structure,   independent on the BI field,
$(\Gamma_{(0)}')^2=1 , tr \,
\Gamma_{(0)}'=0$).
 All dependence on BI field ${\cal F} = ``\tan "Y$ is in the exponent
\cite{BKOP}.
The matrix
 $\Gamma_{11}$ in IIA and  $\sigma_3\otimes \ 1$ in IIB theory
anticommute with $\Gamma_{(0)}'$ and with $\Gamma$. Therefore in the basis
where  $\Gamma_{11}$
and $\sigma_3\otimes \ 1$ are
diagonal, $\Gamma_{(0)}'$ and $\Gamma$ are  off-diagonal. We  introduce
a $16\times 16$-dimensional matrix $\hat  \gamma$ which does not depend on BI
field.
\begin{equation}
\Gamma_{(0)}'  = \pmatrix{0&
\hat  \gamma \cr
\cr               \hat \gamma^{-1} &0}\,  \ ,  \qquad   \Gamma = \pmatrix{0&
\hat \gamma e^{{\hat a}}\cr
\cr                ( \hat \gamma e^{{\hat a}})^{-1}&0}\, ,
\end{equation}
where
\begin{equation}
\hat a=\cases{+{1\over2} Y_{jk} \gamma^{jk} \qquad\qquad &{\rm
IIA\, ,}\cr
-{1\over2} Y_{jk} \gamma^{jk}\qquad\qquad &{\rm IIB\, .}}
\end{equation}

The fact that $\Gamma$ is off-diagonal and that the matrix $\gamma e^{{\hat
a}}$
is invertible is quite important and the significance of this for covariant
quantization of D-branes was already
discussed in \cite{Ag1,BKOP,K97}.
In particular this allows us to consider only irreducible $\kappa$-symmetry
transformations by imposing a Lorentz covariant
condition on $ \bar \kappa$ of the form
\begin{eqnarray}
\bar \kappa_1  &=& 0  \qquad  \bar \kappa_2   \neq  0 \qquad {\rm IIA}
\\
\bar \kappa_2 &=& 0  \qquad   \bar \kappa_1  \neq  0 \qquad {\rm IIB} \ .
\end{eqnarray}
In this way we have an irreducible  16-dimensional $\kappa$-symmetry since
the matrix $\hat \gamma e^{\hat a}$ is invertible, acting as
\begin{eqnarray}
\delta \bar \theta_1 &=& \bar \kappa_2 \hat \gamma e^{{\hat a}}  \qquad
\delta \bar \theta_2 = \bar \kappa_2 \hskip 2.2 cm  \delta X^m = - \bar
\kappa_2\Gamma^m \theta_2
 \hskip 2.2 cm  {\rm IIA} \label{kappaA}\\
\delta \bar \theta_1 &=& \bar \kappa_1 \hskip 1.5  cm
\delta \bar \theta_2 = \bar \kappa_1 (\hat \gamma e^{\hat a} )^{-1} \qquad
\delta X^m = - \bar
\kappa_1\Gamma^m \theta_1
\hskip 2.2  cm  {\rm IIB} \ .  \label{kappaB}
\end{eqnarray}

Recently a covariant gauge  fixing $\kappa$-symmetry  of D-branes has been
discovered \cite{Ag1}. The fermionic gauge is of the form
\begin{eqnarray}
\theta_2 &=& 0 \qquad {\rm IIA}  \qquad \theta_1 \equiv \lambda
\label{thetasA}\\
\theta_1 &=& 0  \qquad {\rm IIB}  \qquad \theta_2 \equiv \lambda \ .
\label{thetasB}
\end{eqnarray}
Note that our choice of irreducible $\kappa$-symmetry (which is not unique) was
made here with the purpose to explicitly eliminate $\theta_2 $ ($\theta_1$) in
IIA (IIB) case using  $\delta \bar \theta_2 = \bar \kappa_2
$ ( $\delta \bar \theta_1 = \bar \kappa_1$).
The gauge-fixed action has one particularly useful property: the Wess-Zumino
term vanishes in this gauge \cite{Ag1}. We are left with the reparametrization
invariant action:
\begin{equation}
S_{\kappa-{\rm fixed}}= - \int d^{p + 1} \sigma \sqrt{- { {\rm det}}
(G_{\mu\nu} + {\cal F}_{\mu\nu})} \ ,
\label{gaugefixed}\end{equation}
\begin{equation}
G_{\mu\nu} = \eta_{mn} \Pi_\mu^m \Pi_\nu^n \ , \qquad \Pi_\mu^m = \partial_\mu
X^m - \bar\lambda \Gamma^m \partial_\mu \lambda \ ,
\end{equation}
\begin{equation}
{\cal F}_{\mu\nu} = [\partial_\mu A_\nu - \bar \lambda \Gamma_m
\partial_{\mu}\lambda
\left(\partial_{\nu} X^m - {1\over 2} \bar
\lambda \Gamma^m \partial_{\nu}\lambda \right)] - (\mu \leftrightarrow \nu) \ .
\end{equation}

This action (\ref{gaugefixed}) has a local reparametrization symmetry and a
32-component global supersymmetry. The form of the action is such that it can
be brought to the form close to the one discovered by Volkov-Akulov \cite{VA}.
Consider for example a D-9-brane in a static gauge \cite{Ag1}  $X^{\mu} =
\sigma^{\mu}$  :
\begin{equation}
S_{ g.f.}= - \int d^{10} \sigma \sqrt{- { {\rm det}}
(G_{\mu\nu} + {\cal F}_{\mu\nu})} \ ,
\label{gaugefixed}\end{equation}
where
\begin{equation}
G_{\mu\nu}= e_ {\mu}{}^m  e_ {\nu}{}^n \eta^{mn}
\end{equation}
and
\begin{equation}
e_ {\mu}{}^m= \delta_ {\mu}{}^m  - \bar\lambda \Gamma^m \partial_\mu \lambda \
{}.
\end{equation}
If we introduce the 1-forms depending on fermion fields $\lambda  (\sigma) $
\begin{equation}
e^m = d \sigma^{\mu} e_ {\mu}{}^m[\lambda (\sigma)] = d \sigma^m +\bar\lambda
\Gamma^m d \lambda
\end{equation}
we can rewrite the supersymmetric Born-Infeld 9-brane action as

\begin{equation}
S= \int e^{m_0} \wedge e^{m_1} \wedge \cdots  \wedge e^{m_9} \; \sqrt{-  {\rm
det}
(\eta_{mn}  + {\cal F}_{mn})}
\label{gaugefixed1}\end{equation}
where ${\cal F}_{mn}=  e^ {\mu}{}_m e^ {\nu}{}_n {\cal F}_{\mu\nu}$.
In absence of the two-form ${\cal F}_{mn}$ the supersymmetric Born-Infeld
action is reduced to geometric action of Volkov-Akulov \cite{VA}, generalized
to d=10:
\begin{equation}
S= \int e^{m_0} \wedge e^{m_1} \wedge \cdots  \wedge e^{m_9} = \int d^{10}
\sigma \;
{\rm det} e [\lambda (\sigma)]
\end{equation}
This action depends only on fermions $\lambda(\sigma)$ and has a non-linearly
realized supersymmetry manifest, since the 1-forms $e^m$ are supersymmetric.
The second supersymmetry of the supersymmetric Born-Infeld 9-brane action
(\ref{gaugefixed1}) is not manifest. It explicit form can be obtained from the
preservation of the gauge-fixing condition for the kappa-symmetry.

The gauge-fixed actions of extended supersymmetric objects in static gauge are
known to lead to complicated non-linear actions. For example, the action of d=2
massive superparticle \cite{AGIT}
\begin{equation}
S=- M \int dt \left( \left [  - (\dot  X^m - \bar \theta \Gamma^m \dot \theta )
 (\dot X^n - \bar \theta \Gamma^n \dot \theta ) \eta_{mn}\right ]^{1/2} - 1 +
\theta \Gamma^3 \dot \theta  \right) \qquad m=0,1.
\end{equation}
quantized in the gauge $\Gamma^3  \theta=\theta$ for $\kappa$-symmetry and
the static gauge  for reparametrization symmetry, $X^0=t$,  gives the kink
effective action \cite{AGIT}
\begin{equation}
S_{g.f}= -M \int dt \left( \left [ 1- \dot \phi^2  \right ]^{1/2} - 1 + {i\over
4M} \rho \dot \rho  \right) \ .
\end{equation}
Here the bosonic field is the remaining coordinate of the d=2 superparticle,
$\phi=X^1$.
 The Hamiltonian associated with this action is also non-linear:
\begin{equation}
H =(p^2 +M^2)^{1/2} - M  \ ,
\end{equation}
 $p$ is the momentum canonical conjugate to $X^1$.

 Here we will perform a covariant quantization of the D-0-brane which is a d=10
generalization of the d=2 massive superparticle \cite{AGIT}. Instead of the
static gauge, which belongs to a class of canonical gauges with the
non-propagating ghosts, we will use a covariant gauge for reparametrization
symmetry.
Consider the $\kappa$-symmetric action of a D-0-brane
\cite{Ce1,Ag1,Be1,Ho1,Hull}. D-0-brane action does
not
have Born-Infeld field since there is no place for an antisymmetric tensor of
rank 2 in one-dimensional theory.
 The D-p-brane action  for $p=0$ case reduces to
\begin{equation}
S = -T \left (  \int d\tau \sqrt { - G_{\tau\tau}}
  +  \int  \bar \theta \Gamma^{11} \dot \theta \right) \ .
\label{0action}\end{equation}
This action can be derived from the action of the massless 11-dimensional
superparticle \cite{Be1}.
\begin{equation}
S = \int d\tau \sqrt {-g_{\tau\tau}}  g^{\tau\tau} \left( \dot X^{\hat m}  -
\bar \theta \Gamma^{\hat m} \dot \theta \right)^2 \ , \qquad \hat m =
 0,1,\cdots , 8,9,10.
\label{11}\end{equation}
We may solve  equation of motion for $X^{ \hat {10}}$    as ${\bf P}_{ \hat
{10}} =Z$, where $Z$ is a constant, and use   $ \Gamma^{11} =
\Gamma^{\hat {10}}
$. From this one can deduce a
 first order action
\begin{equation}
S = \int d\tau \left( {\bf P}_{m} (\dot X^m - \bar \theta \Gamma^m \dot \theta)
+ {1\over 2} V (
{\bf P}^2 + Z^2) - Z \bar \theta \Gamma^{11} \dot \theta + \bar \chi_1
d_2\right) \ .
\label{first}\end{equation}
We will show now that the D-0-brane action can be obtained from this one upon
solving equations of motion for ${\bf P}_{m}$, $V, $  $\chi_1$, and $d_2$.
Here  $V(\tau) $ is a Lagrange multiplier, $Z=T$  is some constant parameter in
front of the WZ term and  ${\bf P}^2 \equiv {\bf P}^{m}\eta_{mn} {\bf P}^{n}$.
 The chiral spinors $\chi_1$ and $d_2$ are auxiliary fields. They are
introduced to close the gauge symmetry algebra off shell. To verify that this
first
order action is one of the D-p-brane family actions given in (\ref{action}) we
can use equations of motion for ${\bf P}_{m}$
\begin{equation}
{\bf P}_{m}= -{1\over V} (\dot X^m - \bar \theta \Gamma^m \dot \theta) \ ,
\end{equation}
and for the auxiliary fields $ \chi_1=0$ and $  d_2=0$. The action
(\ref{first})
becomes
\begin{equation}
S = \int d\tau \left( -{1\over 2 V} (\dot X^m - \bar \theta \Gamma^m \dot
\theta)^2 +{1\over 2} V
 Z^2 - Z \bar \theta \Gamma^{11} \dot \theta \right) \ .
\label{inter}\end{equation}
Equation of motion for $V$ is
\begin{equation}
 V^2=- {1\over Z^2 } (\dot X^m - \bar \theta \Gamma^m \dot \theta)^2 \ ,
\end{equation}
and we can insert $ V= -{1\over Z }\sqrt { - (\dot X^m - \bar \theta \Gamma^m
\dot \theta)^2}
$ back into the action (\ref{inter}) and get
\begin{equation}
S = - Z \left(  \int d\tau \sqrt { - (\dot X^m - \bar \theta \Gamma^m \dot
\theta)^2}
  + Z \bar \theta \Gamma^{11} \dot \theta \right) \ .
\end{equation}
This is the action (\ref{action}) for D-0-brane at $T=Z$ as given in
(\ref{0action}).

The action  (\ref{first}) is invariant under the 16-dimensional irreducible
$\kappa$-symmetry and under the
reparametrization symmetry. The gauge symmetries are (we denote $\Gamma^m {\bf
P}_{m}= \sl$):
\begin{eqnarray}
\delta \bar \theta&=& \bar  \kappa_2 ( \Gamma^{11} Z + \sl) \ , \\
\delta X^m &=& -\eta {\bf P}^{m}  - \delta \bar \theta  \Gamma^m \theta - \bar
\kappa_2 \Gamma^m d \ , \\
\delta V &=& \dot \eta + 4 \bar \kappa_2 \dot \theta + 2 \bar \chi_1 \kappa_2\
, \\
\delta \bar \chi &=& \bar \kappa_2   \dot  \sl \ ,\\
\delta d &=&[ {\bf P}^2 + Z^2] \kappa_2 \ .
\end{eqnarray}
Here $\eta(\tau) $ is the time reparametrization gauge parameter and
$\kappa_2(\tau)
=
{1\over 2} (1- \Gamma^{11} ) \kappa (\tau) $ is the 16-dimensional parameter of
$\kappa$-symmetry. The gauge symmetries form a closed algebra
\begin{equation}
[\delta({\kappa_2}) ,\delta({\kappa'_2}) ] = \delta ({\eta = 2 \bar \kappa_2
\sl \kappa'_2} ) \ .
\end{equation}

To bring the theory to the canonical form we introduce  canonical momenta to
$\theta$ and to $V$ and find, excluding auxiliary fields
\begin{equation}
L =  {\bf P}_{m} \dot X^m+ P_V \dot V + \bar P_{\theta } \dot \theta  + {1\over
2} V ({\bf P}^2
 + Z^2)
  +   P_V \varphi + \left (\bar P_{\theta }  + \bar \theta  (\sl + Z
\Gamma^{11})\right ) \psi \ .
\end{equation}

We have  primary constraints $\bar \Phi  \equiv  \bar P_{\theta }  + \bar
\theta  (\sl + Z \Gamma^{11}) \approx 0$
and $P_V=0$. The Poisson brackets for 32 fermionic constraints are
\begin{equation}
\{ \Phi  ,  \Phi  \}=  2 C (\sl  + \Gamma_{11} Z ) \ .
\end{equation}
We also have to require that the constraints are consistent with the time
evolution $\{P_V, H\}=0$. This generates a secondary constraint
\begin{equation}
t=  {\bf P}^2 + Z^2 \ .
\end{equation}
Thus the Hamiltonian is weakly zero and any physical state of the system
satisfying the reparamet-\\
rization constraint is a BPS state $M=|Z|$ since
\begin{equation}
 {\bf P}^2 + Z^2 | \Psi \rangle  = 0 \qquad  \Longrightarrow  \qquad Z^2   |
\Psi \rangle =-  {\bf P}^2  | \Psi \rangle = M^2  |  \Psi  \rangle \ .
\end{equation}
 The $32\times 32$ -dimensional matrix $C (\sl + \Gamma_{11} Z) $ is not
invertible since it squares to zero when the reparametrization constraint is
imposed. This is a reminder of the fact that D-0-brane is a d=11 massless
superparticle.
The 32 dimensional fermionic constraint has a 16-dimensional part which forms a
first class constraint and another 16-dimensional part which forms a second
class constraint.
We notice that the Poisson brackets reproduce the $d=10$, $N=2$ algebra with
the
central charge which can also be understood as $d=11$, $N=1$ supersymmetry
algebra with the
constant value of ${\bf P}_{11}=Z$.

We proceed with the quantization and gauge-fix
$\kappa$-symmetry covariantly   by taking $\theta_2=0, \theta_1 \equiv \lambda
$
and find
\begin{equation}
L^\kappa_{g.f.}= {\bf P}_{m} (\dot X^m - \bar \lambda  \Gamma^m \dot \lambda )
+ {1\over 2} V (
{\bf P}^2 + Z^2) \ .
\end{equation}
The 16-dimensional fermionic constraint
\begin{equation}
 \bar \Phi_{\lambda }  \equiv  (\bar P_{\lambda }  + \bar \lambda  \sl )\approx
0
\end{equation}
forms the Poisson bracket
\begin{equation}
\{  \Phi^{\alpha }_{\lambda }  ,  \Phi ^{\beta }_{\lambda } \}=   2  (\sl
C)^{\alpha \beta}   \label{Pois} \ .
 \end{equation}
The matrix $ \sl C
$ is perfectly invertible as long as the central charge $Z$ is not vanishing.
The inverse to (\ref{Pois}) is
\begin{equation}
\{  \Phi^{\alpha }  ,  \Phi ^{\beta } \}^{-1}  \mid_{t=0} \;
= [2  ( \sl C)^{\alpha \beta}   ]^{-1} = {(C \sl )_{\alpha \beta}
 \over 2  {\bf P}^2} \ .
  \end{equation}

This proves that the fermionic constraints are second class and that the
fermionic part of the Lagrangian
\begin{equation}
- \bar \lambda  \sl \dot \lambda \equiv - i \lambda^\alpha \Phi_{\alpha \beta}
\dot \lambda^\beta \ ,  \qquad \Phi_{\alpha \beta} =-i (C\sl )_{\alpha \beta} \
,
\end{equation}
is not degenerate in a Lorentz covariant gauge. None of this would be true for
a vanishing central charge. Note that in the rest frame ${\bf P}_{0}=M , \vec
{\bf P}=0
$, hence
\begin{equation}
 \Phi_{\alpha \beta} = M \delta_{\alpha \beta} \ .
\end{equation}
For D-0-brane one can covariantly gauge-fix the reparametrization symmetry by
choosing the $V=1$ gauge and including the anticommuting reparametrization
ghosts
$b,c$. This brings us to the following form of the action:
\begin{equation}
L^{\kappa, \eta} _{g.f.}= {\bf P}_{m} \dot X^m - \bar \lambda  \sl \dot \lambda
 + {1\over 2}  (
{\bf P}^2 + Z^2)
 +  b\dot c \ .
\end{equation}
Now we can define Dirac brackets
\begin{eqnarray}
&& \{  \lambda  ,  \bar \lambda \}^*= \{  \lambda , \bar \Phi \}  \{ \bar \Phi_
,  \Phi  \}^{-1}  \{  \Phi  \ , \bar \lambda \} = {   \sl \over 2 {\bf P}^2} =-
{ \sl \over 2 Z^2} \ .
\end{eqnarray}

The generator of the 32-dimensional supersymmetry is
\begin{equation}
\bar \epsilon Q = \bar \epsilon ( \sl  + \Gamma^{11} Z) \lambda \ .
\end{equation}

It forms the following Dirac bracket

\begin{equation}
[\bar \epsilon Q \ , \bar Q \epsilon' ]^* = \bar \epsilon (\sl +
\Gamma^{11} Z)  {   \sl \over 2 {\bf P}^2} (\sl +
\Gamma^{11} Z)  \epsilon'=
\bar \epsilon \Gamma^{\hat m}  {\bf P}_{\hat m}   \epsilon'   = \bar \epsilon
(\sl +
\Gamma^{11} Z)  \epsilon' \ .
\label{diracsusy}\end{equation}

We can also rewrite it in d=11 Lorentz covariant form
\begin{equation}
[\bar \epsilon Q \ , \bar Q \epsilon' ]^*=\bar \epsilon \Gamma^{\hat m}  {\bf
P}_{\hat m}   \epsilon' = \bar \epsilon \; \hat {\sl } \epsilon'\ ,
 \qquad \hat m =
 0,1,\cdots , 8,9,10, \qquad   Z= {\bf P}_{ \hat {10}}\ ,  \qquad \Gamma^{11} =
\Gamma^{\hat {10}} \ .
\label{diracsusy11}\end{equation}

This Dirac bracket realizes the d=11, N=1 supersymmetry algebra
 or, equivalently, d=10, N=2 supersymmetry algebra with the central charge
$Z$.

One can also  to take into account that the path integral in presence of second
class constraints has an additional term with $\sqrt{{\rm Ber} \{\Phi _\lambda
,
\Phi_\lambda  \}} \sim  \sqrt{ {\rm Ber}\, \Phi_{\alpha \beta} }$~
\cite{Fradkin}. It can be
used to make a change of variables
\begin{equation}
S_\alpha  = \Phi^{1/2}_{\alpha \beta}    \; \lambda^\beta \ .
\end{equation}
The action becomes

\begin{equation}
L= {\bf P}_{m} \dot X^m-  i S_\alpha  \dot S_\alpha   +  b\dot c - H
\end{equation}

\begin{equation}
H= - {1\over 2} (
{\bf P}^2 + Z^2) \ .
\end{equation}

The generators of global supersymmetry commuting with the Hamiltonian take the
form
\begin{equation}
\bar  \epsilon  Q    =   \bar \epsilon ( \sl  + \Gamma^{11} Z)  \Phi^{-1/2} S \
{}.
\end{equation}
Taking into account that $ \{ S _\alpha  ,  S_\beta  \}^* =-{i\over 2}
\delta_{\alpha \beta}$
we have again realized $d=10$, $N=2$  supersymmetry algebra in the form
(\ref{diracsusy}) or  (\ref{diracsusy11}).
The nilpotent BRST operator in this gauge where only reparametrization ghosts
are propagating is given by
\begin{equation}
Q_{BRST} = c H \qquad   H= \{ b,  Q_{BRST}  \}  \qquad (Q_{BRST})^2=0
\end{equation}
and here we used the fact that $\{ b,  c \}=1$. In turn, the Hamiltonian (and
therefore the BRST operator) can be constructed from supersymmetry generators
as follows:
\begin{equation}
H = {1\over 2}  \{ Q _\alpha  ,  Q_\beta  \}^*  \{ \bar Q ^\alpha  ,  \bar
Q^\beta  \}^* =  {1\over 2}  ( C \hat \sl
)_{\alpha \beta} ( \hat \sl
C)^{\alpha \beta} =- {1\over 2}  \hat \sl^2 = - {1\over 2}  (\sl^2 + Z^2)
\end{equation}

The terms with anticommuting fields $S_\alpha $ can be rewritten in a form
where it is clear that they can be interpreted as   world-line spinors,
 \begin{equation}
L= {\bf P}_{m}  \partial_0 X^m +  \bar S_\alpha  \rho^0 \partial_0  S_\alpha
+  b\dot c - H \ .
\end{equation}
Here $\bar S_\alpha =  i S_\alpha  \rho^0$ and $(\rho^0)^2 =-1$, $\rho^0=i$
being a 1-dimensional matrix.

Thus, we have the original 10 coordinates $X^m$ and their conjugate momenta
${\bf P}_{m}$, and a pair of reparametrization ghosts. There are also 16
anticommuting world-line spinors $S$, describing 8 fermionic degrees of
freedom. The Hamiltonian  is quadratic.
The  ground state with   $M^2=Z^2$ is the state with the minimal value of the
Hamiltonian.
Thus for the D-superparticle one can see that the condition for the covariant
quantization is satisfied in the presence of a central charge which makes the
mass of a physical state non-vanishing. The global supersymmetry algebra is
realized in a covariant way, as different from the light-cone gauge.

Thus we have found that covariantly quantized D-0-brane has a quadratic action
with the physical state being the BPS state $M=|Z|$.
The resulting supersymmetry generator is $d=10$ Lorentz covariant and the Dirac
bracket of the quantized theory form $d=10$, $N=2$ supersymmetry algebra with
a central charge.

\end{document}